\documentclass[11pt]{article}

\usepackage[preprint]{acl}

\usepackage{times}
\usepackage{pgfplots}
\usepackage{latexsym}
\usepackage{hyperref}       
\usepackage{url}            
\usepackage{booktabs}       
\usepackage{amsfonts}       
\usepackage{nicefrac}       
\usepackage{microtype}      
\usepackage{xcolor}         
\usepackage{graphicx}
\usepackage{multirow}
\usepackage{tabularx}
\usepackage[T1]{fontenc}

\usepackage[utf8]{inputenc}

\usepackage{microtype}

\usepackage{inconsolata}

\usepackage{graphicx}

\title{AppAgent v2: Advanced Agent for Flexible Mobile Interactions}

\author{
\bfseries
Yanda Li$^{1,2}$\thanks{This work was completed by Yanda during an internship at Tencent GY Lab.},\hspace{1em}
Chi Zhang$^{2,4}$\thanks{Project Leader},\hspace{1em}
Wenjia Jiang$^4$,\hspace{1em}
Wanqi Yang$^{1}$,\hspace{1em}
Bin Fu$^2$, \\
\bfseries
Pei Cheng$^2$,\hspace{1em}
Xin Chen$^2$,\hspace{1em}
Ling Chen$^{1}$,\hspace{1em}
Yunchao Wei$^{3}$
\\[0.1125cm]
\normalsize 
$^1$ University of Technology Sydney\hspace{1.5em}
$^2$ Tencent\hspace{1.5em}
$^3$ Beijing Jiaotong University\hspace{1.5em}
$^4$ Westlake University
\\
\normalsize
$^1$ \texttt{Yanda.Li@student.uts.edu.au, ling.chen@uts.edu.au}\\
\normalsize
$^3$ \texttt{wychao1987@gmail.com} $^4$ \texttt{chizhang@westlake.edu.cn}
}

\begin{document}
\maketitle
\begin{abstract}
  With the advancement of Multimodal Large Language Models (MLLM), LLM-driven visual agents are increasingly impacting software interfaces, particularly those with graphical user interfaces. This work introduces a novel LLM-based multimodal agent framework for mobile devices. This framework, capable of navigating mobile devices, emulates human-like interactions. Our agent constructs a flexible action space that enhances adaptability across various applications including parser, text and vision descriptions. The agent operates through two main phases: exploration and deployment. During the exploration phase, functionalities of user interface elements are documented either through agent-driven or manual explorations into a customized structured knowledge base. In the deployment phase, RAG technology enables efficient retrieval and update from this knowledge base, thereby empowering the agent to perform tasks effectively and accurately. This includes performing complex, multi-step operations across various applications, thereby demonstrating the framework's adaptability and precision in handling customized task workflows. Our experimental results across various benchmarks demonstrate the framework's superior performance, confirming its effectiveness in real-world scenarios. Our code will be open source soon.
\end{abstract}

\section{Introduction}
Large Language Models (LLMs) like ChatGPT~\cite{openai_chatgpt} and GPT-4~\cite{openai2023gpt} have greatly advanced natural language processing, enabling their integration into intelligent agents that revolutionize autonomous decision-making. These agents~\cite{schick2024toolformer,qin2023toolllm}, initially tailored for text-based interactions, exhibit advanced human-like features, including adaptive memories that enhance their environmental engagements and processing capabilities across diverse NLP tasks.

However, real-world applications often require beyond textual processing, necessitating the integration of visual data and other modalities. This requirement exposes shortcomings in traditional text-only agents and highlights the urgent need for advanced multimodal systems. These systems~\cite{gao2023assistgpt,suris2023vipergpt,wu2023visual} are critical in complex environments like mobile and operating system platforms where they need to perform multi-step reasoning, extract and integrate information, and respond adaptively to user inputs. Innovative solutions such as the AppAgent~\cite{yang2023appagent} and MobileAgent~\cite{wang2024mobile} have shown promise by enabling more natural interactions with smartphone applications through human-like interactions.


Despite these advancements, accurately recognizing graphical user interfaces (GUIs) remains a key challenge, impacting the decision-making accuracy of multimodal agents. Previous methods~\cite{liu2024multimodal,wang2024mobile} relying on visual features often face inaccuracies due to limitations in recognition models. Additionally, the dynamic nature of mobile environments, which frequently introduce new features, poses further challenges. Even sophisticated models like GPT-4, while proficient with well-known apps, struggle with lesser-known apps due to unfamiliar visual elements. The rapid updates in app interfaces and functionalities further hinder these models' effectiveness across diverse applications. 

To address this challenge, AppAgent~\cite{yang2023appagent} adopts a human-like approach by automated exploration and watching demos. This strategy allows the agent to store UI element descriptions in a document rather than relying on rigid memorization, thus enhancing decision-making by leveraging contextual understanding. However, AppAgent depends heavily on an off-the-shelf parser to identify UI elements, which restricts the agent's operational flexibility in environments featuring non-standard components such as video players and games. This dependency limits the agent's ability to adapt its actions to unfamiliar or unique interface elements, thereby affecting its overall effectiveness in diverse applications. 

To mitigate these limitations, we propose a novel multimodal agent framework designed to adapt to the dynamic mobile environment and diverse applications. We develop an extensive action space enabling the agent to interact with a wide variety of elements. This includes not only those elements that can be parsed using a standard parser but also elements and text identified through OCR and detection tools.
Unlike previous work that relied solely on ID matching from parser to retrieve information, our approach incorporates multiple forms of element data. To facilitate access diverse elements, we have designed a structured storage system to construct a knowledge base. Each element within the knowledge base can store different attribute information such as parser details, textual content, and visual descriptions. This system is tailored to organize and store element information in a manner that supports quick retrieval and effective utilization, significantly boosting the agent’s ability to perform in novel scenarios.



Following previous work~\cite{yang2023appagent}, our agent operates in two distinct phases: exploration and deployment. In the exploration phase, our agent autonomously analyzes and documents the functionality of unknown UI elements and applications, tailored to specific task types. This proactive documentation allows the agent to build a robust knowledge base of UI layouts and operations, vital for handling tasks in unfamiliar environments. During this phase, we also incorporate a reflection module, which serves to validate the documented functionalities based on iterative assessments, ensuring the accuracy and reliability of the information stored. In the deployment phase, the agent leverages RAG technology~\cite{lewis2020retrieval} to dynamically access and update its knowledge base with relevant document content based on real-time interactions, significantly enhancing its capability to adapt to novel scenarios. This framework not only streamlines the learning process but also enhances the agent's decision-making capabilities by providing a deeper understanding of each application’s functionality.

We validated our agent's effectiveness through tests on three distinct benchmarks, encompassing tasks across numerous applications. Quantitative results and user studies demonstrate the superiority and robustness of our approach. 
In summary, this paper makes the following contributions:

\begin{itemize}

\item We introduce a multimodal agent framework that combines parser with visual features to construct a flexible action space, enhancing interaction with GUI and improving adaptability to new environmental tasks.


\item We develop a new structured storage format that, coupled with RAG technology, allows for adaptive, real-time updates and access to the knowledge base, enhancing the agent's adaptability and decision-making precision.

\item We conduct extensive empirical testing, demonstrating the agent's effectiveness across a variety of smartphone applications, validating its adaptability, user-friendliness, and efficiency in real-world scenarios.
\end{itemize}
\section{Method}
\begin{figure*}[t]
\centering
\includegraphics[width=1.0\textwidth]{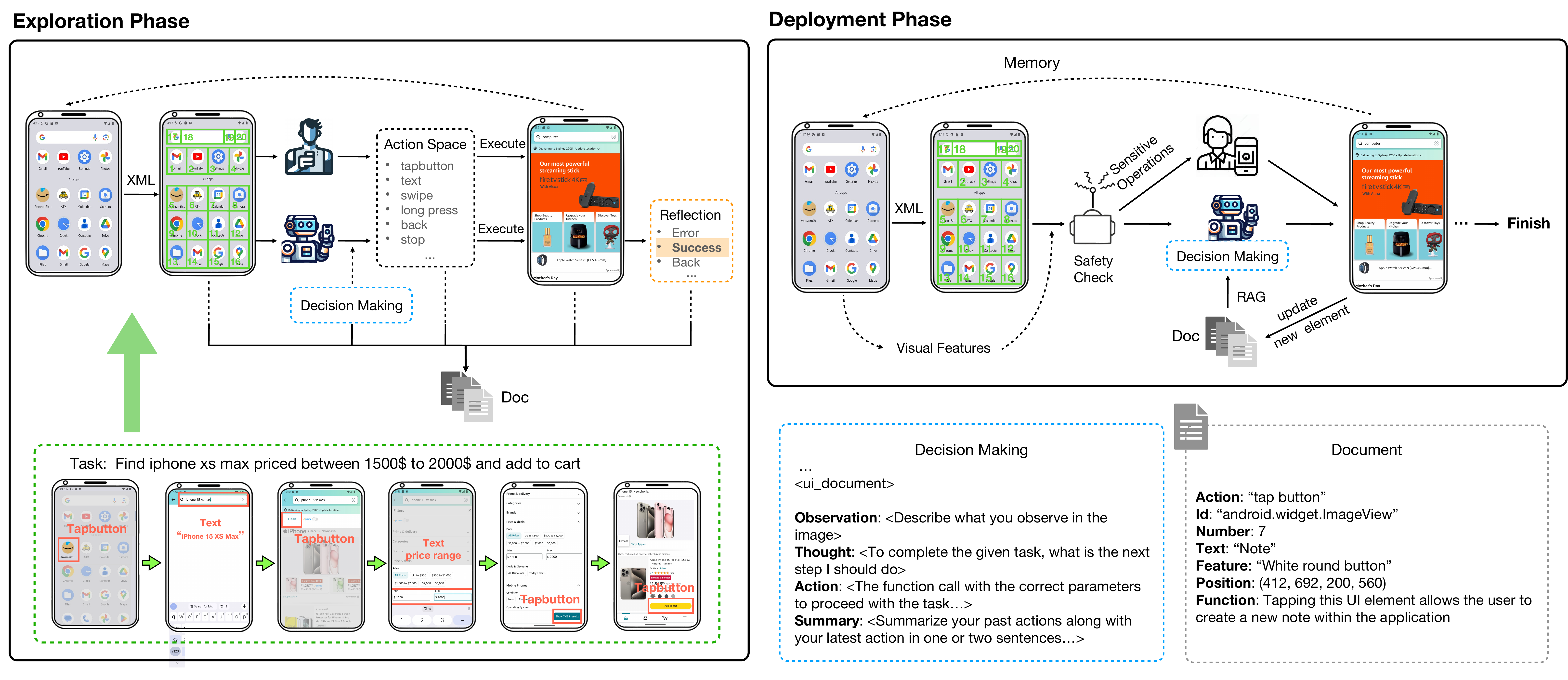}
\caption{Overview of our agent pipeline. Exploration module takes agent-driven or manual exploration collects element information into a document. Deployment phase takes RAG to retrieve and update the document in real time, thereby rapidly preparing to execute tasks} 
\label{fig:pipe}
\vspace{-1.5em}
\end{figure*}

In this section, we provide a detailed description of our multimodal agent framework as Figure~\ref{fig:pipe}, which is structured into two primary phases: exploration and deployment. At each round, the agent analyzes the current GUI with task requirements, generating observations, thoughts, actions, and summaries. The summary, serving as memory, is carried over to the next execution prompt, ensuring continuity throughout the task execution process. 

\subsection{Agent Framework}
Our multimodal agent framework is implemented on the Android 15 environment using the Android Studio emulator. The agent interacts with the mobile phone by invoking commands through the \texttt{AndroidController}. This interaction process is based on analyzing the current GUI interface's structured data parsing information, combined with OCR and detection models to extract detailed information from screenshots. The data extracted includes Android ID, numerical labels marked on the screenshots, features of the elements, texts, and the coordinates of the UI elements. This setup allows the agent to perform efficiently within a dynamic mobile environment, integrating advanced recognition capabilities with intelligent decision-making processes based on the interpreted data from the user interface.

\subsection{Agent Interactions}
During both the exploration and execution phases, the agent interacts with the mobile phone, translating human commands or outputs from LLMs into instructions that the Android system can recognize and execute. We detail these commands as follows:

\begin{enumerate}
  \item \textbf{TapButton:} Initiates tap action on user interface element. This can be specified either by entering the element's number identifier in the screenshot or by describing its visual features.
  
  \item \textbf{Text:} Simulates typing by entering a string of text into the designated area. 

  \item \textbf{LongPress:} Applies a prolonged press on a specified element area. 

  \item \textbf{Swipe:} Executes a swipe action in a specified direction on an element. This can be used for scrolling pages vertically or horizontally.

  \item \textbf{Back:} Simulates the device's back button to return to the previous UI state.

  \item \textbf{Home:} Commands the agent to return to the main screen. This is crucial for agent to re-execute the tasks and cross-apps tasks.

  \item \textbf{Wait:} Pauses the operation to allow the system to catch up, refresh the screen snapshot.

  \item \textbf{Stop:} Signals the completion of tasks and ends the current operation. 

\end{enumerate}

Once these commands are transformed into corresponding instructions, they are executed by the Android system through the \texttt{AndroidController}. This setup ensures precise command execution, allowing the agent to perform tasks efficiently within the Android environment. More details about action space are displayed in Appendix.

\subsection{Exploration Phase}
The exploration phase is aimed at analyzing the GUI in relation to the current task. It involves identifying and documenting the functions of UI elements through two alternative methods: agent-driven and manual exploration. All prompts used are displayed in Appendix.

\subsubsection{Agent-Driven exploration}
This method starts with the agent analyzing the current UI interface to identify elements requiring interaction and to determine the specific actions needed. Once these elements and actions are pinpointed, the agent executes the planned actions.
Following the execution of action, the agent takes screenshots before and after the interaction to compare and analyze the changes. This comparison allows the agent to record the operational functions of the UI elements and assess the effectiveness of each action taken. 

Afterwards, the agent enters reflection phase. If the agent determines that the executed action is completely irrelevant to the task, it performs a return operation. The irrelevant action is recorded in a \textit{useless\_list} and is fed back into the LLM. If the results of the actions align with the intended user task and prove effective, the relevant UI information is documented and continued to explore. 

This reflection ensures that only actions that align with the user's task are considered effective and documented for future retrieval. This method not only enhances the quality of the knowledge base but also refines the agent's strategy in real-time, ensuring that subsequent actions are more likely to contribute effectively to task completion.

\subsubsection{Manual Exploration}
This method is introduced to overcome the limitations encountered during agent-driven exploration, such as the LLM's erroneous judgments due to its incomplete understanding of certain apps and UI elements. Manual exploration allow GPT-4 to observe manual operations, compare screenshots before-and-after operations similar to agent-driven, gaining a clearer understanding of new UI elements and task workflows. The exploration is enhanced with advanced OCR and detection models, providing comprehensive UI analysis based on human interactions. Humans guide the sequence of actions and conclude the process, thereby streamlining the operational workflow and accelerating the learning process. Importantly, just like in automatic exploration, the information regarding UI elements and their functionalities observed during manual exploration is meticulously documented. This manual exploration ensures that the agent can overcome shortcomings of the automated processes by incorporating sophisticated understanding and adjustments that only human insight can provide.

\subsection{Development Phase}
During the deployment phase, the agent can utilizes the knowledge acquired to perform user tasks effectively. Initially, the agent fetches the current GUI information and traverses the elements using Self-query retriever for document retrieval. The self-query retriever converts document content into embeddings, stored in a vector store, from which it retrieves the most pertinent document based on resource IDs or OCR-derived information.

The agent then integrates this document into the prompt for agent, analyzing the current GUI screenshot, document content, and specific task requirements to make informed decisions and execute actions based on the positional information of UI elements. \textit{Alternatively, the agent can also operate without loading the document, directly handling the majority of common tasks effectively.} After each action, the agent updates its prompts with historical information and action outcomes, thereby enhancing its memory and improving decision-making for subsequent steps. 

The process continues until the agent determines that the task has been completed, at which point it exits the current process and reports task completion. This structured approach ensures that actions are executed precisely and efficiently, leveraging the detailed knowledge base created during the exploration phase to optimize performance and user satisfaction.

\subsection{Document Generation}
This document serves as a specialized knowledge base, meticulously designed to store comprehensive information about UI elements collected during the exploration phase. The database includes various data for each UI element such as Android ID, visible labels, text content, visual features (e.g., color and shape), screen coordinates, and functionalities as interpreted by GPT-4.

To enhance accessibility and utility, we have developed a novel structured storage format suitable for managing diverse element types. This format not only facilitates organized data retrieval but also supports dynamic updates based on real-time interactions during the deployment phase. As the agent operates across various applications, it actively updates the document in response to new UI elements and adapts its strategies accordingly.

This dynamic updating mechanism ensures that the agent remains adaptable and efficient, capable of adjusting its actions based on user requirements and contextual changes. The continual enhancement of the document significantly improves the agent’s understanding and manipulation of application interfaces, leading to more accurate and contextually appropriate interactions. Meanwhile, markedly enhances the user experience and operational efficiency of the agent.



\subsection{Advanced Features}
This subsection highlights the key functionalities that enhance our multimodal agent framework, focusing on visual feature decision-making, safety checks, and cross-app task management. These features collectively improve the agent's safety, versatility, and efficiency, ensuring robust performance in complex and dynamic environments.

\subsubsection{Visual Features Decision-Making}
When the agent confronts scenarios where the desired interactive element is not numerically tagged, and other numerically tagged elements are ineffective for task completion, it automatically transitions to an alternative visual feature UI layout. This process leverages advanced OCR technology~\cite{liao2020real} and detection models~\cite{liu2023grounding} to accurately recognize and annotate text and icons within the interface. By numerically annotating these elements using established methodologies, the agent is equipped to make informed decisions based on the newly adapted UI screenshot. This capability is crucial for handling icons in previously unknown scenarios, ensuring that the agent can navigate and interact with various UI elements effectively, regardless of prior exposure. This dynamic decision-making process significantly enhances the agent's ability to adapt to new environments and execute tasks with higher precision and reliability.

\subsubsection{Safety Check}
In modern LLMs and agent systems, safety is crucial, particularly in automated processes that can lead to privacy breaches. To tackle this issue, we implemented a safety check during the deployment phase. The agent reviews the current UI screenshot, and if the next steps involve sensitive actions like account passwords, payment or other privacy-related concerns, it will switch to manual mode so the user can handle these operations personally. For privacy, the agent will not retain any information from this process. Once the user completes the sensitive task and inputs "finish," the agent will automatically continue with the deployment phase and carry on with the task until it’s completed.
The safety check offers several key advantages. It ensures that sensitive tasks remain secure by involving human judgment and minimizes the risk of data leakage. Furthermore, it increases user trust in the system, providing assurance that private information is handled carefully, while still enabling the agent to effectively complete its assigned tasks.

\subsubsection{Cross-Apps Task}
In addition to its core functionalities, our framework is capable of handling complex tasks that span multiple applications. This ability allows the agent to perform tasks that require interactions across different interfaces. When engaging in such cross-app tasks, the agent evaluates its progress based on memories and the specific task requirements. It determines whether the actions within one application have been completed before navigating back to the application interface. Subsequently, it assesses the next set of commands and continues executing tasks in another application. This capability is particularly valuable for tasks that involve gathering and processing information from various sources or coordinating actions between different apps. 
\section{Experiments}
In this section, we will conduct a comprehensive evaluation with our agent framework. The experiments were conducted on the Android platform to maintain consistency and simplify validation. We utilized the Android Studio emulator for the experiments, which included comprehensive testing on the public benchmarks and qualitative results. This dual approach allowed us to benchmark our agent against standardized criteria while also gaining deeper insights into its real-world performance on mobile applications and environments.

\subsection{Quantitative Results}
In this section, we present a comprehensive evaluation of our agent using three distinct benchmarks: DroidTask~\cite{wen2024autodroid}, AppAgent~\cite{yang2023appagent}, and Mobile-Eval~\cite{wang2024mobile}.  We begin with DroidTask to test complex task performance, comparing against AppAgent for different exploration methods, and conclude with Mobile-Eval to assess comprehensive capabilities. Results in the ensuing sections demonstrate the superiority of our approach in varied application scenarios.

\subsubsection{DroidTask}
In this study, we employ the DroidTask dataset~\cite{wen2024autodroid}, an Android Task Automation benchmark suite designed to evaluate the capabilities of mobile task automation systems. DroidTask consists of 158 high-level tasks derived from 13 popular applications.
We conducted our experiments using the DroidTask dataset. Due to variations in the app versions and device models used during evaluation, the specific workflows for implementing functionalities in the apps may differ. Consequently, we employ the "Completion Rate" as our evaluation metric, similar to~\cite{wen2024autodroid}. The Completion Rate is defined as the probability of accurately completing all the actions in a given sequence, which gauges the agent's ability to consistently and successfully execute a task.

AutoDroid incorporates a memory mechanism, analogous to the document in our agent. We compared the performance of AutoDroid with and without the memory component to our agent, which is deployed directly without document. We employed the robust LLM GPT-4 as the baseline, and compared our method against the LLM-Framework and two versions of AutoDroid, as illustrated in Table~\ref{tab:droidtask}. The results reveal that our agent, even without exploration stage, not only significantly outperformed GPT-4 but also surpassed AutoDroid when it is augmented with memory. This finding underscores the superiority of our approach in leveraging direct deployment strategies effectively and highlights the robustness of our system in a competitive benchmark environment.

\begin{figure}[!t]
\centering
\begin{tikzpicture}
\small
\begin{axis}[
    width=\linewidth, 
    height=5.5cm, 
    ybar,
    symbolic x coords={LLM-F, AD-w/o M, AD-w M, Ours},
    xtick=data,
    nodes near coords,
    ylabel={Completion Rate (\%)},
    title={Completion Rate Comparison on GPT-4},
    enlarge x limits=0.15,
]
\addplot coordinates {(LLM-F,31.6) (AD-w/o M,63.5) (AD-w M,71.3) (Ours, 77.8)};
\end{axis}
\end{tikzpicture}
\caption{Performance Comparison between AutoDroid and ours on DroidTask with GPT-4} 
\label{tab:droidtask} 
\vspace{-1.5em}
\end{figure}

\subsubsection{AppAgent}
AppAgent~\cite{yang2023appagent} has introduced a benchmark that spans ten commonly used applications with diverse functionalities, including Twitter, Telegram, Temu, among others. We compare our agent against AppAgent on this benchmark to assess our agent's adaptability across various functions and interfaces. The primary evaluation metric is the success rate, which reflects the proportion of tasks that the agent successfully completes within an application. The results are detailed in Table~\ref{tab:appagent}. The results of our agent-driven exploration are comparable to those obtained from AppAgent with watching demos. After integrating the documents generated through manual exploration, our agent's performance improved significantly, underscoring the effectiveness of our exploration phase.

\begin{table}[ht]
\centering
\caption{Quantitavie results between AppAgent and ours.}
\resizebox{\columnwidth}{!}{
\begin{tabular}{@{}lccc@{}}
\toprule
Method        & Document         & Action Space & SR (\%) \\ \midrule
GPT4 (Baseline) & None             & Raw          & 2.2     \\
                & None             & AppAgent     & 48.9    \\
                \hline
AppAgent       & Auto. Exploration & AppAgent     & 73.3    \\
                & Watching Demos   & AppAgent     & 84.4    \\
                \hline
Ours            &  Agent-Driven    & Ours   & 84.4  
\\              &  Manual          & Ours   & \textbf{93.3}                
\\ \bottomrule

\end{tabular}
\label{tab:appagent}
}
\vspace{-1.5em}
\end{table}

\subsubsection{Mobile-Eval}
We evaluated our agent on the Mobile-Eval benchmark. Mobile-Eval is a comprehensive benchmark introduced for mobile agents, containing 10 commonly used mobile apps to test agent performance across different tasks. Mobile-Eval assesses the following metrics:

\begin{itemize}
    \item \textbf{Success (Su):} Marks an instruction as successful if the agent completes it entirely.
    \item \textbf{Process Score (PS):} Evaluates step accuracy by calculating the ratio of correct steps to total steps.
    \item \textbf{Relative Efficiency (RE):} Compares the steps taken by the agent to human performance to measure efficiency.
    \item \textbf{Completion Rate (CR):} Measures the proportion of steps the agent completes compared to a human's total steps.
\end{itemize}


\begin{table*}[ht]
\centering
\resizebox{0.8\textwidth}{!}{%
\begin{tabular}{@{}llllllllllllll@{}}
\toprule
\multicolumn{1}{c}{\textbf{App}} & \multicolumn{4}{c}{\textbf{INSTRUCTION 1}} & \multicolumn{4}{c}{\textbf{INSTRUCTION 2}} & \multicolumn{4}{c}{\textbf{INSTRUCTION 3}} \\ \cmidrule(l){2-13} 
\multicolumn{1}{c}{} & \textbf{SU} & \textbf{PS} & \textbf{RE} & \textbf{CR} & \textbf{SU} & \textbf{PS} & \textbf{RE} & \textbf{CR} & \textbf{SU} & \textbf{PS} & \textbf{RE} & \textbf{CR} \\ \midrule
\textbf{MobileAgent} & & & & & & & & & & & & \\
Alibaba.com & \checkmark & 0.75 & 4/3 & 100\% & \texttimes & 0.39 & 13/8 & 62.5\% & \checkmark & 0.9 & 10/9 & 100\% \\
Amazon Music & \checkmark & 0.44 & 9/5 & 80\% & \checkmark & 0.75 & 8/6 & 100\% & \texttimes & 0.50 & 12/3 & 66.7\% \\
Chrome & \checkmark & 1.00 & 4/4 & 100\% & \checkmark & 0.80 & 5/4 & 100\% & \checkmark & 0.43 & 8/5 & 100\% \\
Gmail & \checkmark & 1.00 & 4/4 & 100\% & \texttimes & 0.56 & 9/8 & 37.5\% & \texttimes & 0.56 & 9/8 & 37.5\% \\
Google Maps & \checkmark & 1.00 & 5/5 & 100\% & \checkmark & 1.00 & 6/6 & 100\% & \checkmark & 1.00 & 6/6 & 100\% \\
Google Play & \checkmark & 1.00 & 3/3 & 100\% & \checkmark & 0.50 & 10/4 & 100\% & \checkmark & 1.00 & 3/3 & 100\% \\
Notes & \texttimes & 0.57 & 7/4 & 100\% & \checkmark & 0.67 & 6/4 & 100\% & \checkmark & 1.00 & 5/5 & 100\% \\
Settings & \checkmark & 1.00 & 4/4 & 100\% & \checkmark & 1.00 & 4/4 & 100\% & \checkmark & 1.00 & 4/4 & 100\% \\
TikTok & \checkmark & 1.00 & 4/4 & 100\% & \checkmark & 1.00 & 10/10 & 100\% & \checkmark & 1.00 & 7/7 & 100\% \\
YouTube & \checkmark & 1.00 & 4/4 & 100\% & \checkmark & 1.00 & 9/9 & 100\% & \checkmark & 1.00 & 7/7 & 100\% \\
Multi-App & \checkmark & 1.00 & 6/6 & 100\% & \checkmark & 1.00 & 10/10 & 100\% & \checkmark & 1.00 & 10/10 & 100\% \\
\midrule
Avg & 0.91 & 0.89 & 4.9/4.2 & 98.2\% & 0.82 & 0.77 & 7.9/6.3 & 90.9\% & 0.82 & 0.84 & 7.5/6.2 & 91.3\% \\
\midrule
\textbf{Ours} & & & & & & & & & & & & \\
Alibaba.com & \checkmark & 1.00 & 3/3 & 100\% & \checkmark & 0.89 & 9/8 & 100\% & \checkmark & 0.82 & 11/9 & 100\% \\
Amazon Music & \checkmark & 1.00 & 5/5 & 100\% & \checkmark & 1.00 & 6/6 & 100\% & \checkmark & 1.00 & 3/3 & 100\% \\
Chrome & \checkmark & 1.00 & 4/4 & 100\% & \checkmark & 0.80 & 5/4 & 100\% & \checkmark &  1.00 & 5/5 & 100\% \\
Gmail & \checkmark & 1.00 & 4/4 & 100\% & \checkmark & 0.80 & 5/4 & 100\% & \checkmark & 1.00  & 8/8 & 100\%  \\
Google Maps & \checkmark & 1.00 & 5/5 & 100\% & \checkmark & 1.00 & 6/6 & 100\% & \checkmark & 1.00 & 6/6 & 100\% \\
Google Play & \checkmark & 1.00 & 4/4 & 100\% & \checkmark & 1.00 & 4/4 & 100\% & \checkmark & 1.00 & 4/4 & 100\% \\
Notes & \checkmark & 0.80 & 5/4 & 100\% & \checkmark & 0.80 & 5/4 & 100\% & \checkmark & 0.80 & 5/4 & 100\% \\
Settings & \checkmark & 1.00 & 4/4 & 100\% & \checkmark & 1.00 & 4/4 & 100\% & \checkmark & 1.00 & 4/4 & 100\% \\
TikTok & \checkmark & 1.00 & 4/4 & 100\% & \checkmark & 1.00 & 10/10 & 100\% & \checkmark & 1.00 & 7/7 & 100\% \\
YouTube & \checkmark & 1.00 & 4/4 & 100\% & \checkmark & 1.00 & 9/9 & 100\% & \checkmark & 1.00 & 7/7 & 100\% \\
Multi-App & \checkmark & 1.00 & 6/6 & 100\% & \checkmark & 0.83 & 12/10 & 100\% & \checkmark & 0.83 & 12/10 & 100\% \\
\midrule
Avg & \textbf{1.00} & \textbf{0.97} & \textbf{4.3}/4.2 & \textbf{100\%} & \textbf{1.00} & \textbf{0.91} & \textbf{6.7}/6.3 & \textbf{100\%} & \textbf{1.00} & \textbf{0.95} & \textbf{6.7}/6.2 & \textbf{100\%}  \\
\bottomrule
\end{tabular}%
}
\caption{Quantitative results of MobileAgent and ours on Mobile-Eval.}
\label{mobile-eval}
\end{table*}

We compared our agent's performance against the original Mobile-Agent benchmark scores and human performance, as shown in Table~\ref{mobile-eval}. Without integrating the documentation and solely relying on the deployment phase, we achieved the results outlined below The upper table shows the results for Mobile-Agent, and the lower table presents results for our agent. Our agent excelled in completing each task, achieving a 100\% success rate across all instructions in the 10 task categories. The average PS score across three instruction sets exceeded 90\%, indicating that our agent efficiently and accurately completed tasks with minimal errors. This demonstrates its ability to closely emulate human behavior and execute specified tasks effectively on various general apps.

\subsection{User study}
\begin{figure*}[t]
\centering
\includegraphics[width=1.0\textwidth]{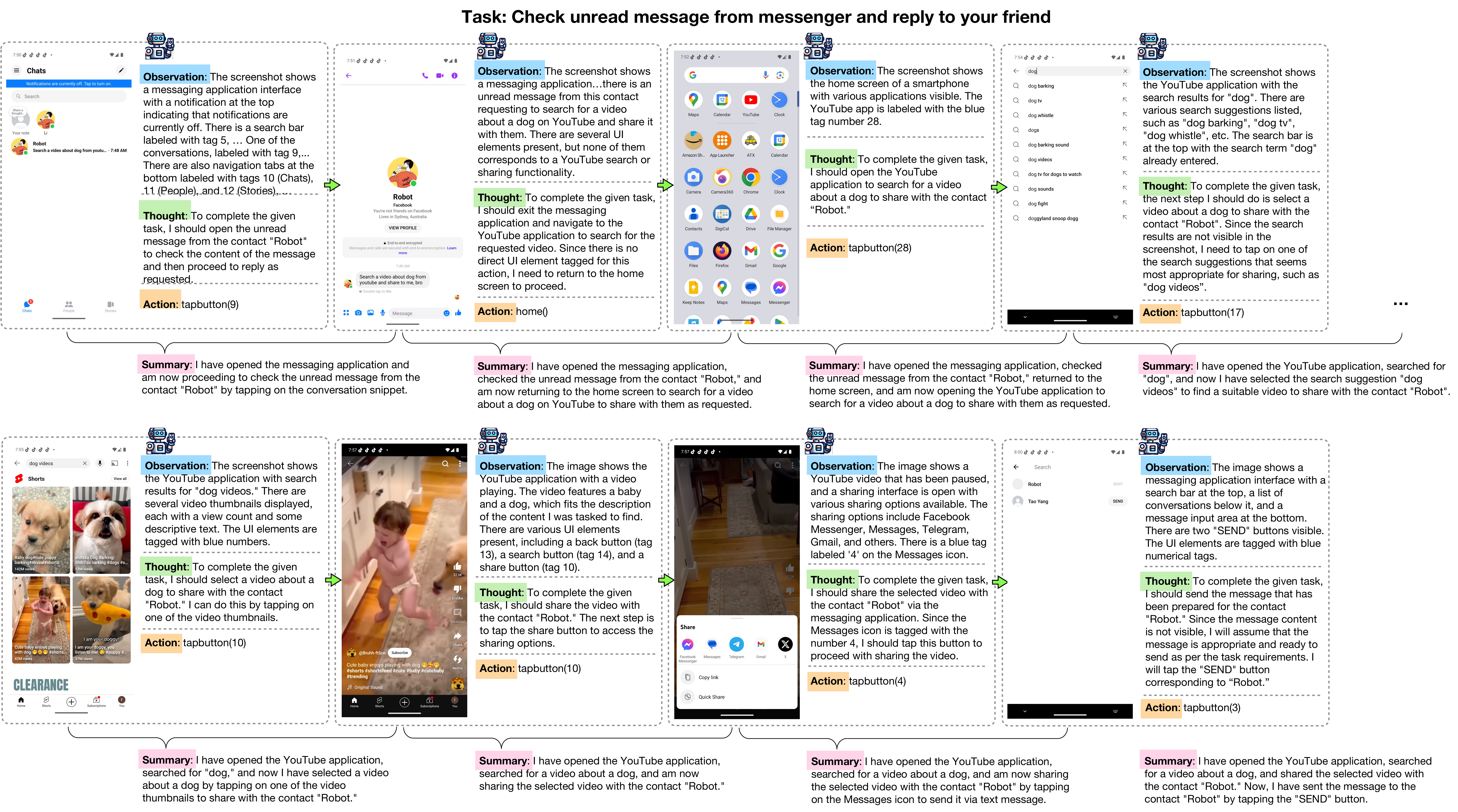}
\caption{Qualitative results of a cross-app task. } 
\label{fig:smart}
\vspace{-1em}
\end{figure*}

To demonstrate our qualitative results, we conducted a user study, as shown in Figure~\ref{fig:smart}. The task involved a series of complex operations, including cross-application activities, long-term multi-step task execution, and multi-step memory storage. To conserve space, we only present the core eight steps of the process here. As can be seen, our agent exhibited outstanding performance in executing complex tasks. More case studies in Appendix.

\subsection{Analysis of UI Interface Parsing}
In our agent, we employ two primary methods for parsing UI interfaces: structured data and visual features. Structured data provides precise and rich information, including details about widget interactivity—such as clickability and scrollability. In this experiment, we utilized XML data parsed from Android systems to enhance our understanding and manipulation of these interactive elements. This method is well-suited for most generic apps and, in conjunction with our agent, can complete the majority of tasks efficiently.

Nevertheless, there are challenges associated with mobile platforms that feature custom-developed apps and icons. Specifically, structured data cannot be parsed for custom icons built on Android, which necessitates the use of visual features for extracting widget information. This approach allows for more accurate recognition of text and icons. However, visual features alone cannot determine the operability of icons without direct interaction, which may lead to redundant operations, such as the agent attempting to interact with non-interactive elements.

Therefore, in our agent, visual feature analysis serves as a secondary operation. It is only employed when the agent determines that no XML-based icons can perform the required task. This strategy enhances the robustness of our agent and improves its transferability to novel apps.

\section{Related works}
\subsection{LLM-based agents}
Agents have rapidly evolved with the advancement of large language models. Models such as MetaGPT~\cite{hong2023metagpt}, HuggingGPT~\cite{shen2024hugginggpt}, and AssistGPT~\cite{gao2023assistgpt}, Seeclick~\cite{cheng2024seeclick} have demonstrated exceptional performance in agent applications, garnering widespread adoption across various domains. Some agents employ large language models such as ChatGPT~\cite{openai_chatgpt} or GPT-4~\cite{openai2023gpt} for task decision-making, achieving notable developments in general domains including music~\cite{huang2024audiogpt,yu2023musicagent}, gaming~\cite{wu2023smartplay, meta2022human}, and autonomous driving~\cite{mao2023gpt,wen2023road,zhou2023vision}. Other agents utilize popular open-source models like LLaMA~\cite{yang2024gpt4tools} and LLaVA~\cite{liu2023llava}. Meanwhile, agents have achieved significant breakthroughs in the multimodal, including video understanding~\cite{yang2024doraemongpt,gao2023assistgpt,wang2023chatvideo}, embodied AI~\cite{yang2023octopus,qin2023mp5}, and visual generation~\cite{chen2023llava_interactive,yang2023mmreact,li2023stablellava}. Additionally, there has been a rise in multi-agent cooperative systems~\cite{qin2023mp5,lee2023explore,long2023discuss} where different agents assume distinct roles. This collaborative approach significantly enhances the capabilities of individual agents, thereby facilitating the achievement of ultimate objectives.

\subsection{Agent for mobile devices}
There are already several agents developed for mobile devices that utilize large language models effectively. DroidBot-GPT~\cite{wen2023droidbot} automates Android app interactions by interpreting app GUI states and actions into natural language prompts, thus facilitating action selection. AppAgent~\cite{yang2023appagent} identifies and enumerates UI components based on XML, subsequently making decisions and executing actions with the aid of GPT-4V. MobileAgent~\cite{wang2024mobile} incorporates visual features, integrating OCR technology and icon detection to enhance UI recognition capabilities.
AutoDroid~\cite{wen2023droidbot} seamlessly combines large language models with dynamic app analysis to optimize mobile task automation efficiently. MobileGPT~\cite{lee2024explore}, an innovative mobile task automator powered by LLMs, is equipped with a human-like app memory system. This system aids in precise task learning and adaptation by structuring procedures into modular sub-tasks, thereby enhancing the performance and flexibility of mobile agents.
\section{Conclusion}
This paper introduces a multimodal agent framework that significantly enhances the interaction capabilities of smartphone applications. Our experiments across various applications demonstrate the framework's ability to improve GUI recognition and task execution, confirming its effectiveness in adapting to diverse application environments.

We integrate parsers with visual features to construct a more flexible action space and develop a newly structured knowledge base for diverse element storage. Through two phases, exploration and deployment, we enable the agent to effectively manage the dynamic nature of mobile interfaces. These capabilities not only align with but also extend the current research on intelligent agents, especially in the contexts of multimodality and mobility.

While building upon existing technologies, our approach contributes incremental advancements in the precision and adaptability of agents operating within complex mobile environments. Future work will focus on enhancing cross-application functionalities and refining decision-making processes to further improve the efficiency and user experience.
\section{Limitations}
Throughout the comprehensive testing process, we identified several limitations of our agent:
Our method relies on the agent's ability to recognize numerical tags on the UI to determine specific UI elements. This approach can lead to confusion when the UI element itself contains numbers. Such errors can be mitigated through preliminary manual exploration and documentation to clarify the context.

When attempting to interact with hidden UI elements, such as accelerating a video by clicking on the screen, the agent lacks the necessary prior knowledge and cannot detect the acceleration button within the current UI. This limitation hampers its ability to perform specific operations. Future work will focus on enhancing UI recognition and incorporating prior knowledge to address these issues effectively.
\section{Ethics Statement}
Our research introduces a novel multimodal agent framework designed to interact seamlessly with smartphone applications, enhancing both user experience and decision-making capabilities. In developing and deploying this technology, we are committed to addressing several key ethical considerations:

Privacy and Data Protection: We ensure strict adherence to global privacy standards, implementing robust data security measures to protect user information.

Reliability and Safety: We implement safety checks to ensure the reliability of our agent, particularly in dynamic environments.

Societal Impact: We consider the broader impacts of our technology, including potential effects on employment and environmental sustainability.

Continuous Monitoring: We commit to continuously monitoring and refining our technology to address emerging challenges and integrate user feedback.

\newpage
\bibliography{custom}

\appendix

\newpage
\section{Prompt Structure Description}
In this section, we describe the main prompts used by our agent, highlighting their structure and purpose across different operational phases. The parts enclosed in bold black angle brackets are parameters that can be replaced during the coding phase, while the red text indicates areas to be filled in by the user, and the blue text represents annotations.

\section{Explanation of DroidTask Results}
In figure~\ref{tab:droidtask}, we present the performance of our agent and AutoDroid on the DroidTask benchmark. The differential in testing environments, AutoDroid’s real device testing on specific Android phone compared to our emulator-based approach, alongside discrepancies in application versions between the two setups, precluded direct execution of some tasks. For a small subset of tasks that could not be completed, we identified alternative testing methods. For instance, whereas our application lacks a date-sorting option for document names, we considered sorting by the initial letter of the document names as an alternative. This adjustment maintains the same procedural flow and steps, albeit with a slightly different selection at the end. Additionally, there are tasks that our application does not support and for which no alternative exists; these cases were treated as error examples. Therefore, under identical conditions, the performance of our agent would be higher than currently observed.

\section{Details of Action Space}
In this section, we provide a detailed description of the usage and parameters for each action space:

\begin{itemize}
    \item \textbf{TapButton(element: int/str)}: Initiates a tap action on a user interface element. For example, \texttt{TapButton(5)} taps the UI element labeled as `5`. \texttt{TapButton('hat')} taps the UI element with text 'hat'.
    \item \textbf{Text(text: str)}: Simulates typing by entering a string of text into a designated input area. For instance, \texttt{Text("Hello, world!")} inputs the string "Hello, world!" into the text field.
    \item \textbf{LongPress(element: int)}: Applies a prolonged press on a specified element. For example, \texttt{LongPress(3)} applies a long press to the element labeled `3`.
    \item \textbf{Swipe(element: int, direction: str, dist: str)}: Executes a swipe action in a specified direction on an element. For instance, \texttt{Swipe(21, "up", "medium")} swipes up on element `21` for a medium distance.
    \item \textbf{Back()}: Simulates the device's back button to return to the previous UI state. Useful for navigating back without specific UI interactions.
    \item \textbf{Home()}: Commands the agent to return to the main screen, essential for resetting the environment or starting new tasks.
    \item \textbf{Wait()}: Pauses the operation for two seconds to allow system processes to complete. 
    \item \textbf{Stop()}: Ends the current operation, signaling the completion of tasks. Useful to terminate processes or to finalize script execution.

\end{itemize}

\section{Case Study}
\begin{figure*}[t]
\centering
\includegraphics[width=1.0\textwidth]{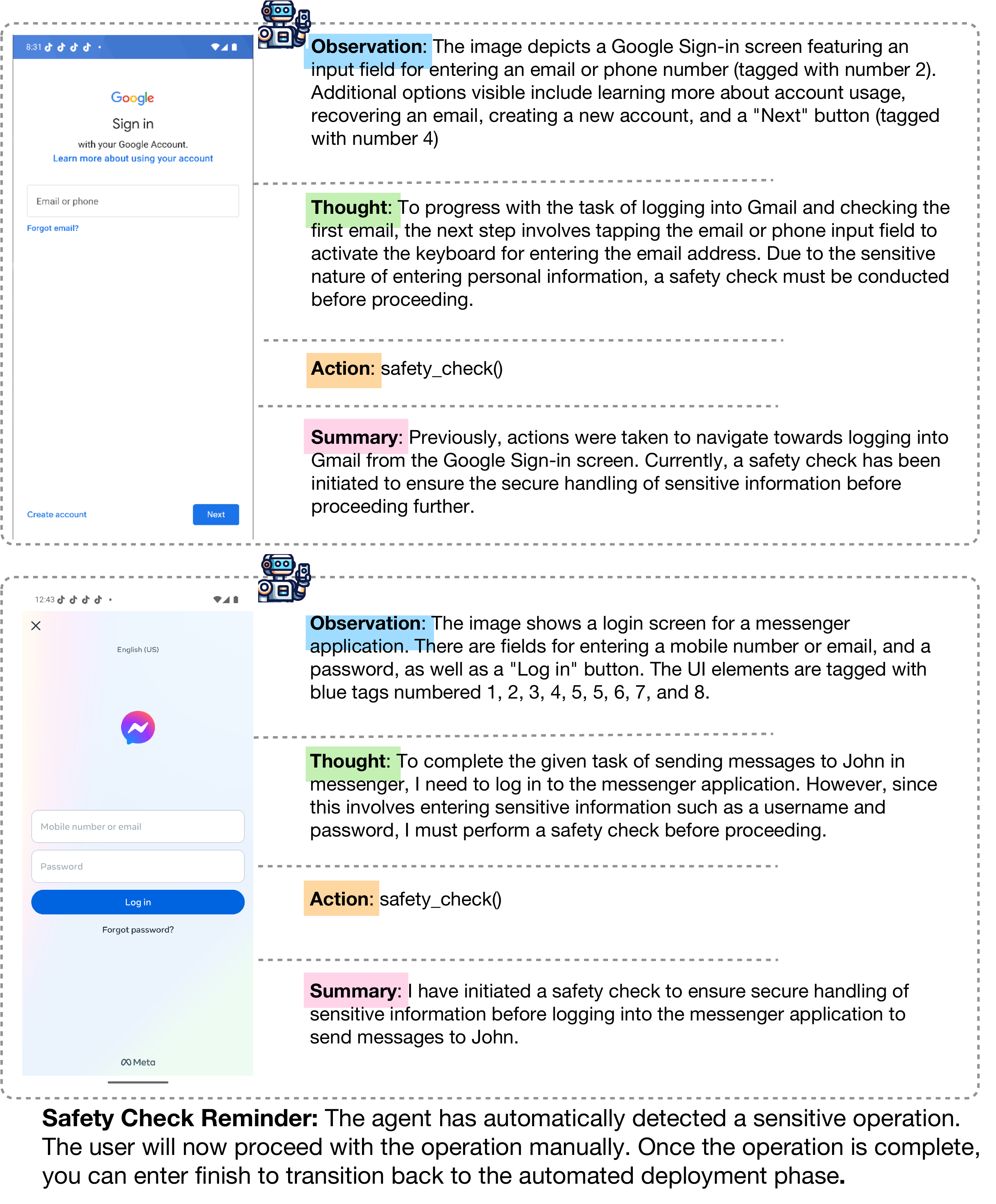}
\caption{Safety check examples.} 
\label{fig:supp1}
\end{figure*}

\begin{figure*}[t]
\centering
\includegraphics[width=1.0\textwidth]{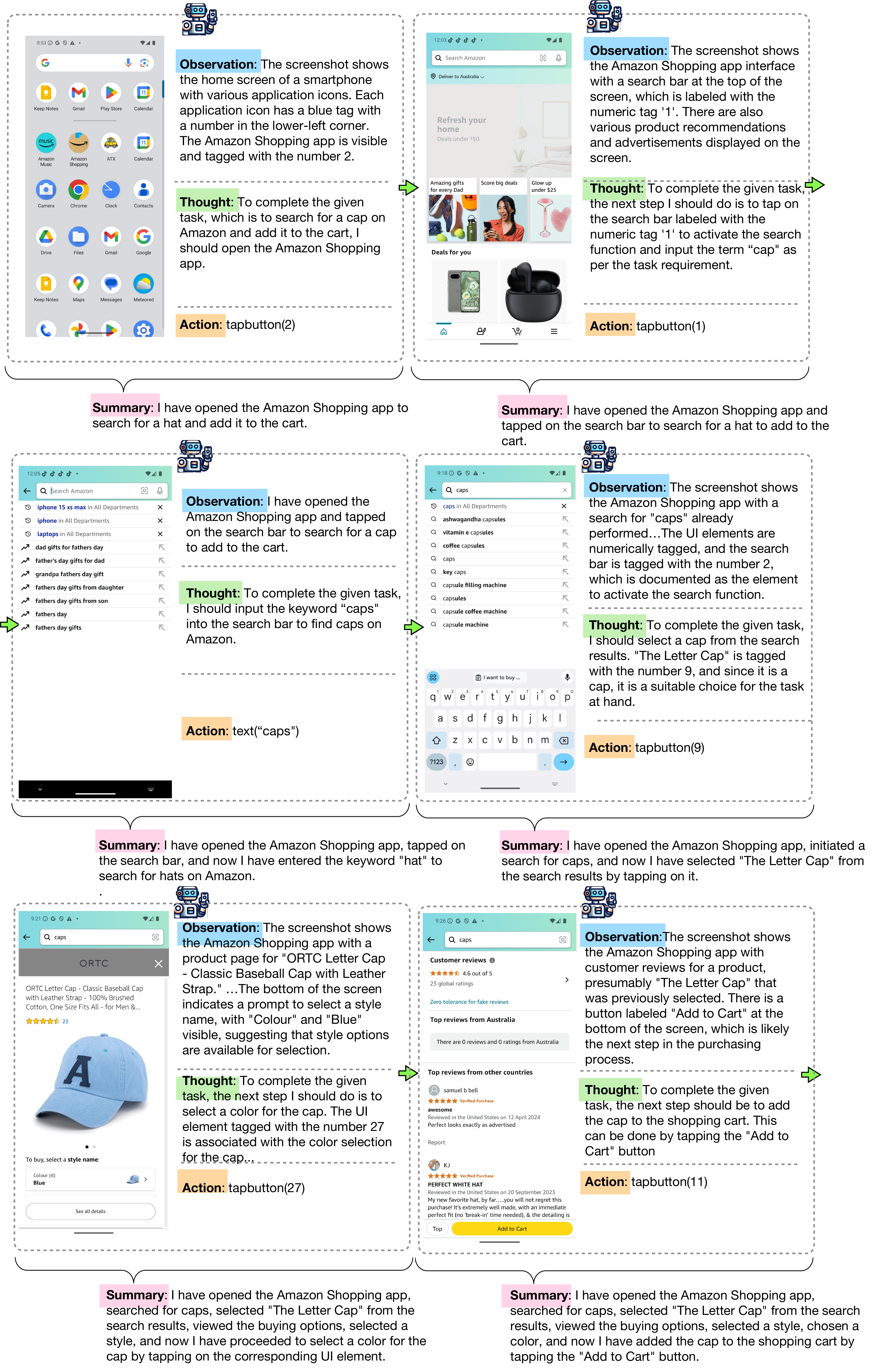}
\caption{Task: search cap in Amazon and add one to cart. } 
\label{fig:supp2}
\end{figure*}

\begin{figure*}[t]
\centering
\includegraphics[width=1.0\textwidth]{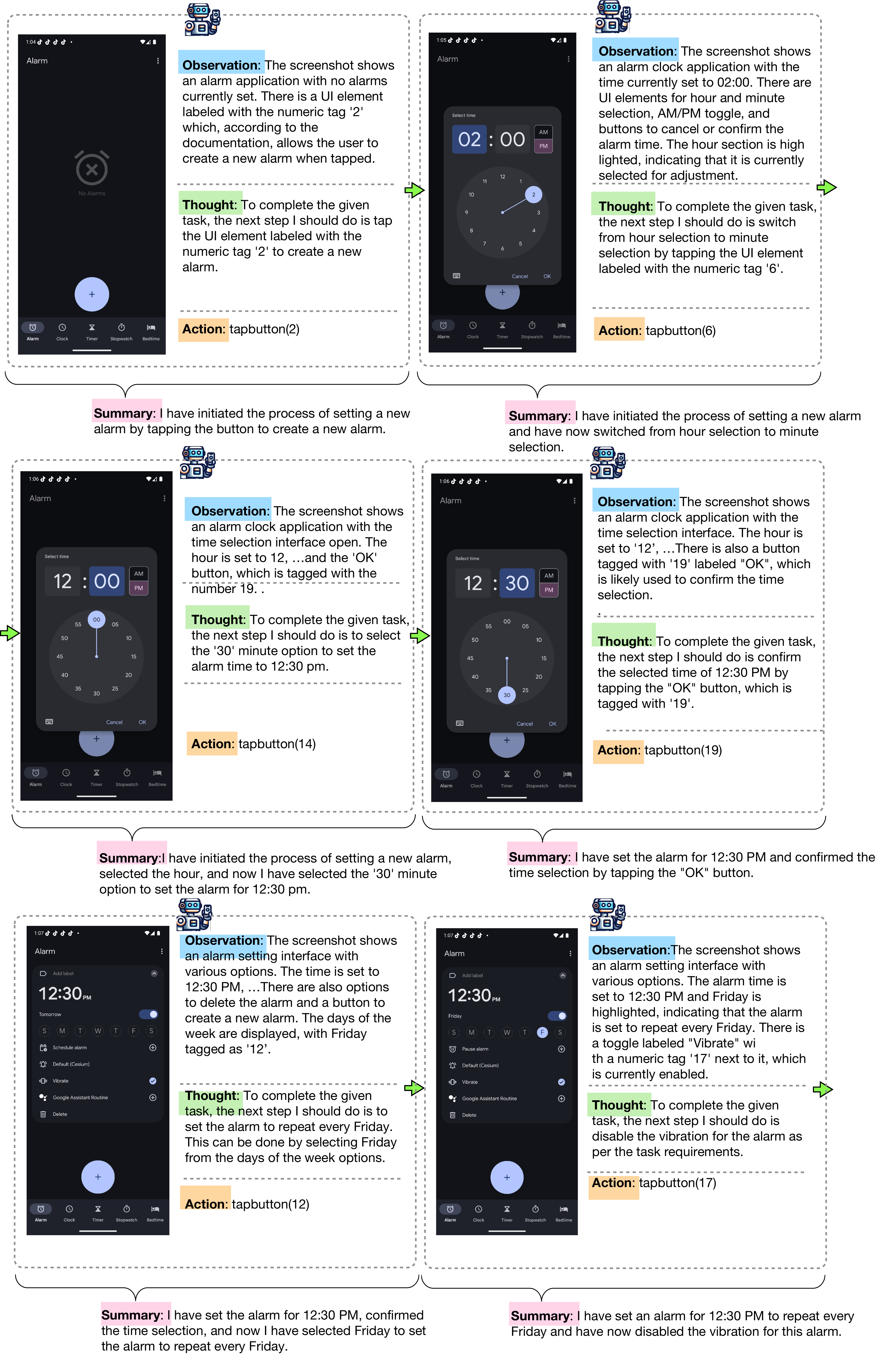}
\caption{Task:set an alarm at 12:30 pm every Friday, disable the vibration. } 
\label{fig:supp3}
\end{figure*}

As illustrated in Figures~\ref{fig:supp1}, ~\ref{fig:supp2} and ~\ref{fig:supp3}, we present several case studies showcasing the qualitative results obtained across diverse applications, tasks, and specialized functionalities. Figure~\ref{fig:supp1} highlights a scenario where our agent triggers a safety check during sensitive operations. Figures~\ref{fig:supp2} and ~\ref{fig:supp3} display the qualitative results of our agent handling multi-step tasks. These examples demonstrate the robustness of our agent, emphasizing its capability to effectively manage a variety of complex scenarios.

\begin{figure*}[t]
\centering
\includegraphics[width=0.8\textwidth]{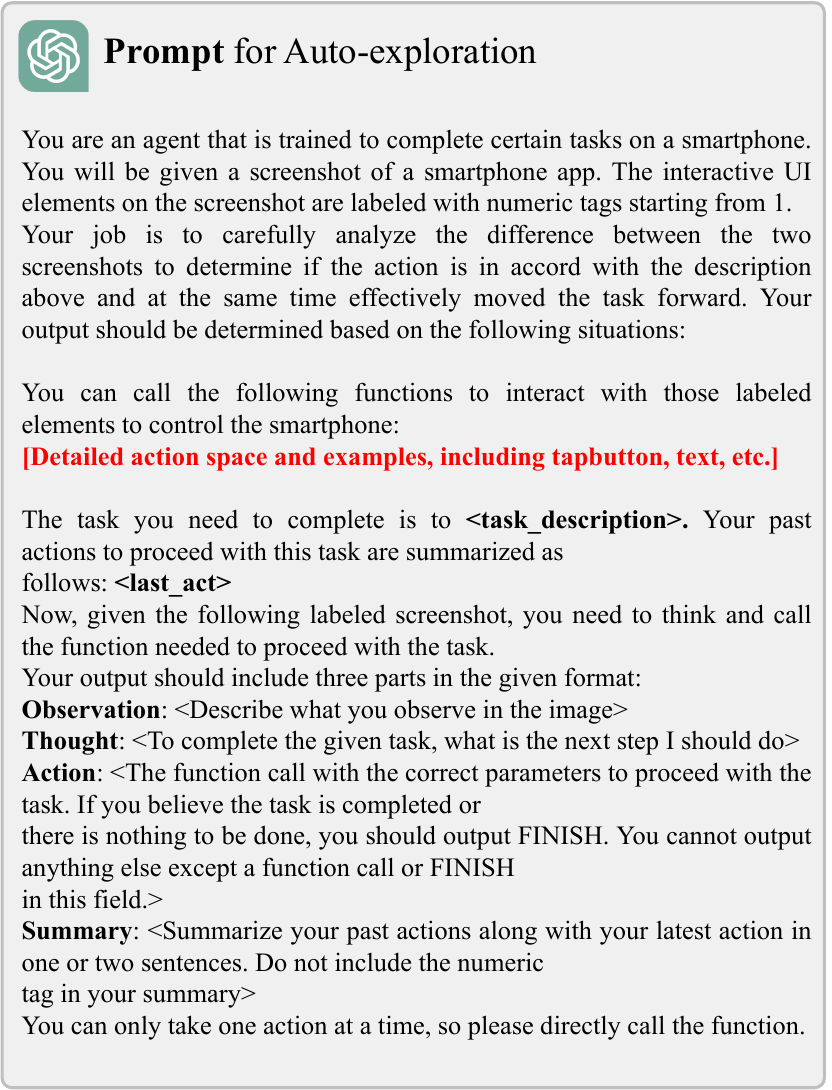}
\caption{Prompt for agent-driven exploration phase. } 
\label{fig:auto}
\end{figure*}

\begin{figure*}[t]
\centering
\includegraphics[width=0.8\textwidth]{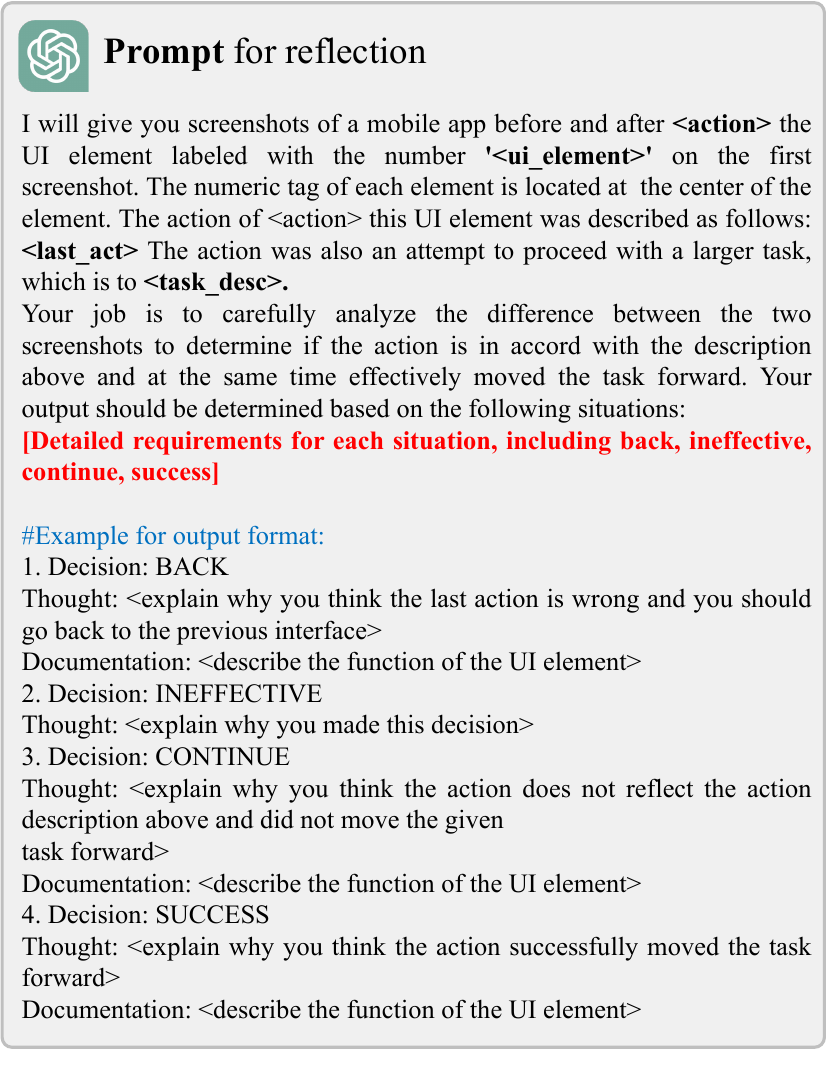}
\caption{Prompt for reflection phase. } 
\label{fig:reflect}
\end{figure*}

\begin{figure*}[t]
\centering
\includegraphics[width=0.8\textwidth]{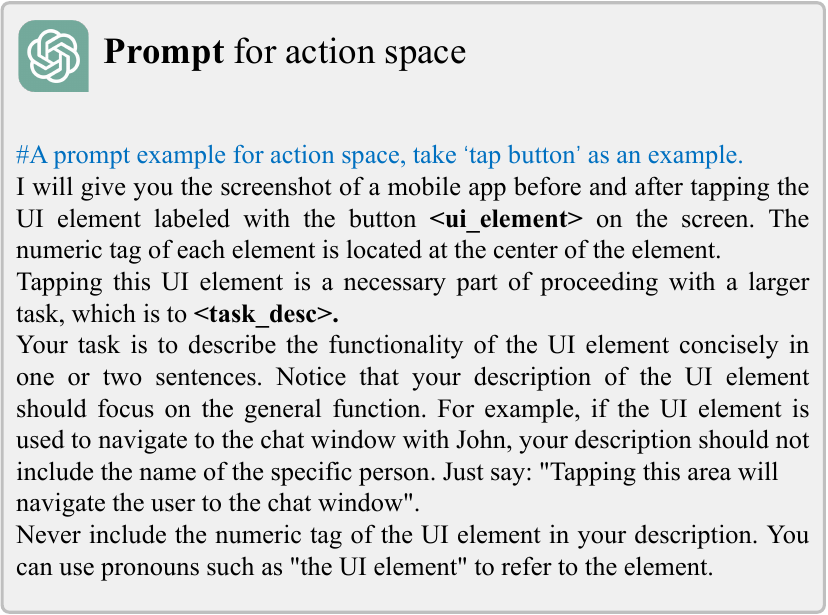}
\caption{Prompt for action space. } 
\label{fig:action}
\end{figure*}

\begin{figure*}[t]
\centering
\includegraphics[width=0.8\textwidth]{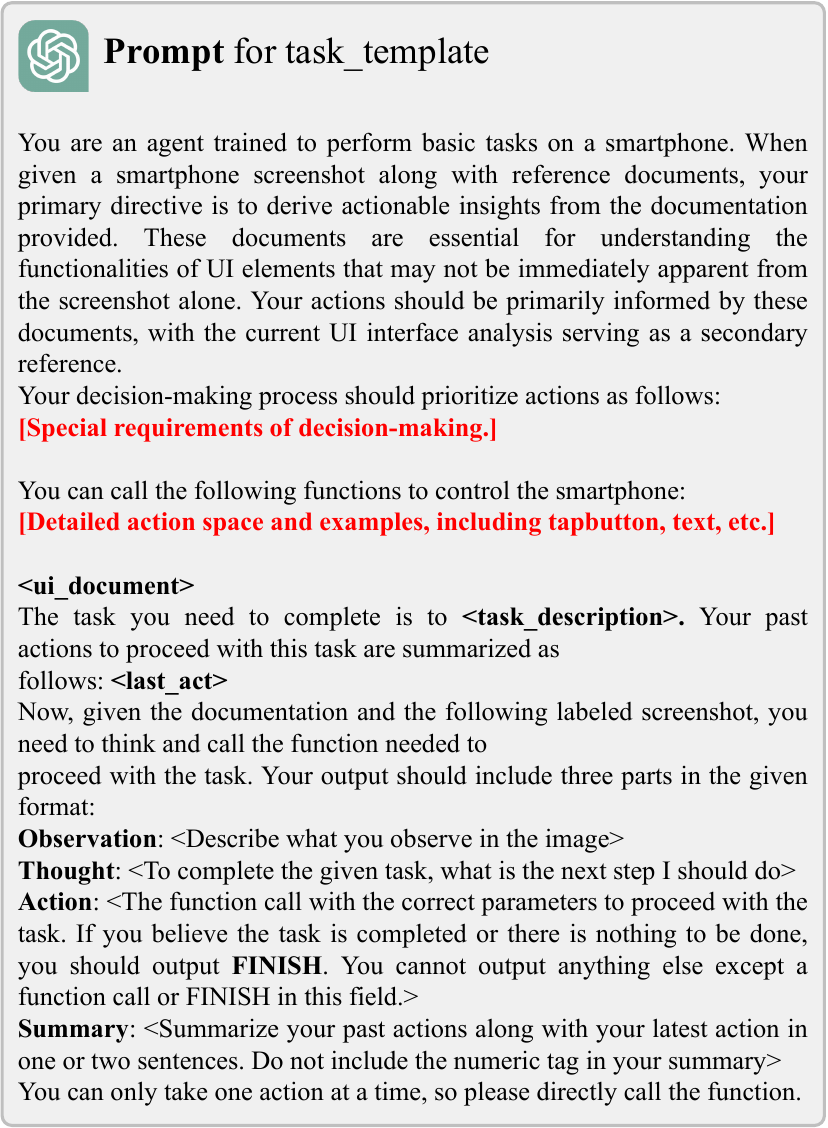}
\caption{Prompt for development phase. } 
\label{fig:task}
\end{figure*}

\label{sec:appendix}

\end{document}